\documentclass{an_art}
\usepackage{graphicx,psfig}
\begin{document}

\renewcommand{\vec}[1]{\mbox{\boldmath $#1$}}
\newcommand{\per}[1]{\mbox{$#1'$}}
\psfigurepath{/turu/rue/drecker/drecker}

%
%

\def\refit{\it}
\def\refem{\bf}

%
%

\def\scr{\scriptsize}            
\def\bl{\par\vskip 12pt\noindent}
\def\bll{\par\vskip 24pt\noindent}
\def\blll{\par\vskip 36pt\noindent}
\def\ul{\underbar} 
\def\ea{et al.}                                         
\def\eg{{\it e.g.}\ }                                    
\def\ie{{\it i.e.}\ }                                   
\def\cf{{\it cf.}\ }
\def\kms{km/s\ }
\def\cms{cm$^2$/s\ }
\def\beg{\begin{equation}}
\def\ende{\end{equation}}

\def\deg{\ifmode^\circ\else$^\circ$\fi}                  
\def\solar{\ifmode_{\mathord\odot}\else$_{\mathord\odot}$\fi} 
\def\gapprox{$_ >\atop{^\sim}$}          
\def\lapprox{$_ <\atop{^\sim}$}          
\def\gsim{\lower.4ex\hbox{$\,\buildrel >\over{\scriptstyle\sim}\,$}}
\def\lsim{\lower.4ex\hbox{$\,\buildrel <\over{\scriptstyle\sim}\,$}}
\def\CaII{Ca\thinspace {\smallrm II}}

\def\onerule{\noalign{\medskip\hrule\medskip}} 
\def\curl{\mathop{\rm curl}\nolimits}
\def\div{\mathop{\rm div}\nolimits}
\def\rot{\mathop{\rm rot}\nolimits}
\def\grad{\mathop{\rm grad}\nolimits}
\def\apj{ApJ\ }
\def\mn{MNRAS\ }
\def\aa{A\&A\ }
\def\aasup{ A\&AS }
\def\csss{ Cool Stars, Stellar Systems, and the Sun}
\def\an{ Astron. Nachr. }
\def\sp{ Solar Phys.}
\def\gafd{Geophys. Astrophys. Fluid Dyn.\ }
\def\ass{ Ap\&SS}
\def\acta{ Acta Astron.}
\def\jfm{ J. Fluid Mech.\ }
\def\ara{ARA\&A}
\def\AIP{Astrophysikalisches Institut Potsdam}
\def\bib{\bibitem{}}

\begin{titlepage}
\setcounter{page}{75}
\headnote{Astron.~Nachr.~ (2001) 75--84}
\makeheadline
\title{Turbulence-driven angular momentum transport in  modulated Kepler flows}
\author{{\sc G. R\"udiger}, Potsdam, Germany\\
\medskip 
{\small Astrophysikalisches Institut Potsdam} \\
\bigskip
{\sc A. Drecker}, Potsdam, Germany\\
\medskip
{\small Astrophysikalisches Institut Potsdam} }
\date{Received 2001 May 17; accepted 2001 May}

\maketitle
\summary
The velocity fluctuations in a spherical shell arising from
sinusoidal  perturbations of a  Keplerian shear flow with a free amplitude 
parameter $\varepsilon$ are studied 
numerically by means of fully 3D nonlinear simulations.
The  investigations are performed at high Reynolds numbers, i.e.   
 3000 $<{\rm Re}<$ 5000.  
We find Taylor-Proudman columns of large eddies parallel to the rotation
axis for sufficiently strong perturbations. 
An instability sets in at critical amplitudes with
$\varepsilon_{\rm crit} \propto {\rm Re}^{-1}$. 
The whole flow turns out to be almost axisymmetric and
nonturbulent exhibiting, however, a very rich radial and
latitudinal structure. 
The Reynolds stress  $\langle u'_r u'_{\phi}\rangle$ is
positive in the entire computational domain,  from its Gaussian
radial profile a positive viscosity-alpha of about $10^{-4}$ is derived.
The kinetic energy of the turbulent state is dominated by the azimuthal component 
$\langle u_\phi'^2\rangle$ whereas the other components are smaller by two 
orders of magnitude. Our simulations reveal, however,  that these structures disappear as soon 
as the perturbations are switched off. 
We did not find an ``effective'' perturbation whose amplitude is such that the 
disturbance is sustained for large times (cf. Dauchot \& Daviaud 1995) which is 
due to the effective violation of the Rayleigh stability criterion. 
The fluctuations rapidly smooth the original profile towards to pure Kepler 
flow which, therefore, proves to be stable in that sense.
END

\keyw{Turbulence theory, nonlinear hydrodynamics, angular momentum transport}
END

\end{titlepage}
\section{Introduction}
The reason for the massive transport of angular momentum which 
enables the formation of the protoplanetary  disk is the key question
in accretion disk physics.
Mostly  this mechanism operates at very high Reynolds numbers
so that the influence of molecular viscosity is assumed to be negligible.
This raises the problem of explaining efficient transport
in  nearly inviscid flows, which are still a challenge
for numerical simulations because of the need for high
numerical resolution due to the rich small-scale
structure.

It is widely accepted
that small-scale turbulence emerging from shear flow instabilities can
lead to enhanced {\em outward} transport -- in opposition to the properties
of convection phenomena in thin Keplerian  disks.
For the latter case several {\em hydrodynamic}  computations have revealed  an 
{\it inwards} transport of angular momentum, i.e. towards the high rotation
rates. Ruden \ea\ (1988) estimated that a modest effective viscosity
will result from the largest convective eddies.
 Ryu \& Goodman (1992) started to find negative
correlations $\langle u_r' u_\phi'\rangle$ for convection in
Keplerian disks, and Cabot \& Pollack (1992) found at low Reynolds
numbers that the Reynolds stress can change sign with increasing
rotation rate. Kley \ea\ (1993) derived from their numerical studies
that large-scale convective motions can lead to an inward flux
of angular momentum.
Negative values for the mentioned  cross correlation also appear in the
simulations of Stone \& Balbus (1996) probing 
the role of vertical convective motions to provide  angular
momentum transport in a Keplerian disk. In R\"udiger \ea\
(2001) it is shown that an anisotropy in the form of dominating
radial velocity fluctuations
\beg
\langle u_r'^2\rangle > \langle u_\phi'^2\rangle
\label{0}
\ende
should be responsible for negative cross correlations and v.v. 
Anisotropic turbulence fields under the influence of global rotation
exhibits extra terms in the cross correlations  
$\langle u_r' u_\phi'\rangle$ which do not
vanish for rigid rotation (`$\Lambda$-effect', see R\"udiger
1989). Its sign depends on  anisotropies  in the turbulence
field such as in (\ref{0}). The question arises whether pure 
Keplerian shear flow instabilities also fit this concept, although
they are known to be linearly stable according to the Rayleigh criterion.

For the solar nebula differential rotation alone has been suggested
by Dubrulle (1993) as a possible explanation for the angular
momentum transfer.  By means of a stability analysis it has been shown
that finite but localized perturbations to the mean
shear flow, which can occur due to pressure fluctuations or material
impact, could lead to an instability independent of 
disk properties like temperature, pressure or density.

Such finite amplitude perturbations for shear flows have been investigated
numerically mainly in plane Couette flow, which is known to be
linearly stable and due to the lack of curvature of the velocity
profile the equations which have to be simulated are
simplified significantly.
Orszag \& Kells (1980) have shown that a transition to turbulence 
at Re $ = 1250$ requires fully 3D disturbances, but 
since they have used a rather low numerical resolution  
	($ 16 \times 33 \times 16 $)  they could not simulate the
fully developed turbulent state. 
Finite-amplitude steady-state solutions have been found by Nagata
(1990) for Re $ = 125$.
Lundbladh \& Johansson (1991) studied
the development of turbulent spots  in plane Couette flow for Reynolds
numbers between $300$ and $1500$ by means of direct numerical
simulations, and actually they found turbulence
to be sustained for sufficiently high Reynolds numbers, Re $ > 375$.
Corresponding experiments have been performed by Dauchot \& Daviaud
(1995) for Re $ < 400$ to
determine the critical amplitude $\varepsilon$ and found a power-law behavior 
\beg 
\varepsilon \sim (\mbox{Re}-\mbox{Re}_{\mbox{\scr NL}})^{-
\alpha}
\ \  {\rm with} \ \  0.3 < \alpha < 0.8, 
\label{1}
\ende
where $\mbox{Re}_{\mbox{\scr NL}}$ is the nonlinear critical Reynolds
number, below which no sustained spots have been observed.
This expression, valid only in the neighborhood of 
$\mbox{Re}_{\mbox{\scr NL}}$,  resembles a transition in terms of 
critical phenomena. However, for large Reynolds numbers an asymptotic
behavior like 
\beg
\varepsilon
 \sim  \mbox{Re} ^{-\beta}
\label{2a}
\ende
holds. The parameter $\beta$ was shown by 
Dubrulle \& Zahn (1991) to be $1/3$ for plane Couette flow.  

As has been pointed out by Balbus \& Hawley (1996) the
interaction between the velocity fluctuations and the background
mean flow differs fundamentally for  cartesian shear-flows
and differential rotation.
They used inviscid simulations of the Euler equations
at moderate resolution of $64^3$, but while turbulence was sustained 
in shear layers they found no numerical evidence that
Keplerian flow at high Reynolds numbers is nonlinearly
unstable. 

For accretion disks this problem now seems to have been solved
by the finding that even a pure Keplerian shear flow
can be unstable when weak magnetic fields are present (Balbus \& Hawley 1991).
This magnetic shear flow instability does only operate in
magnetically well coupled disks if the
magnetic field is neither too weak nor too strong.

In this paper we present results concerning the nonlinear stability
of a perturbed Kepler flow for Reynolds numbers in the
range $3000 < \mbox{Re} < 5000$ obtained by
fully nonlinear simulations of the Navier-Stokes equation. 
First we started with some sinusoidal internal noise perturbing the mean
Keplerian shear flow  and determined the critical amplitude for a given Reynolds
number. In some sense, the modulation of the Kepler flow is used in order to 
simulate force-driven turbulence. 
Afterwards, when the kinetic energy has equilibrated,
the perturbations have been switched off so that a pure
Kepler flow remains. The kinetic energy decays very fast
within a small fraction of a viscous time, thus our
simulations finally confirm that a Kepler flow is nonlinearly
stable even at high Reynolds numbers.
\section{Mathematical formulation}
We focus on the local and global properties
of turbulence arising from finite amplitude perturbations.
We consider a differentially rotating spherical shell of an incompressible
fluid with inner radius  $R_{\mbox{\scr i}}$ and outer radius
$R_{\mbox{\scr o}}$, which enables us to study the large scale
structure as well.

\begin{figure}
\mbox{\psfig{figure=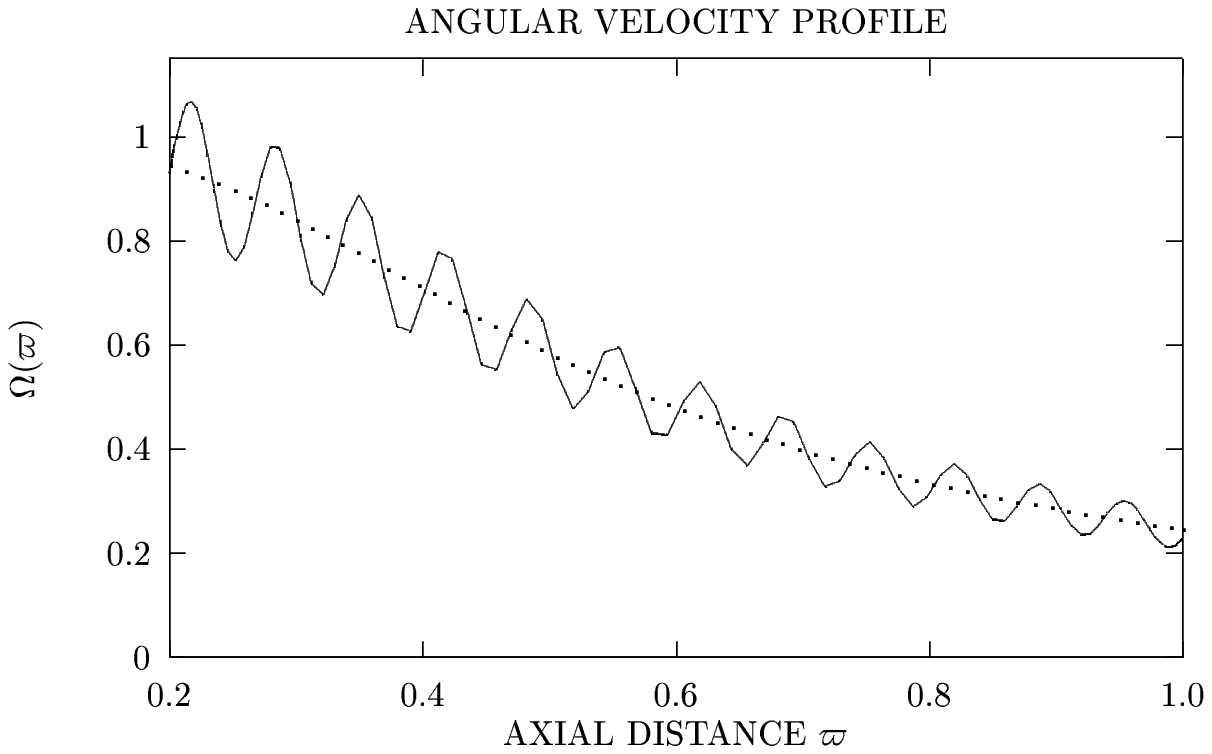,width=8.0cm,height=7.0cm} \hfill
     \psfig{figure=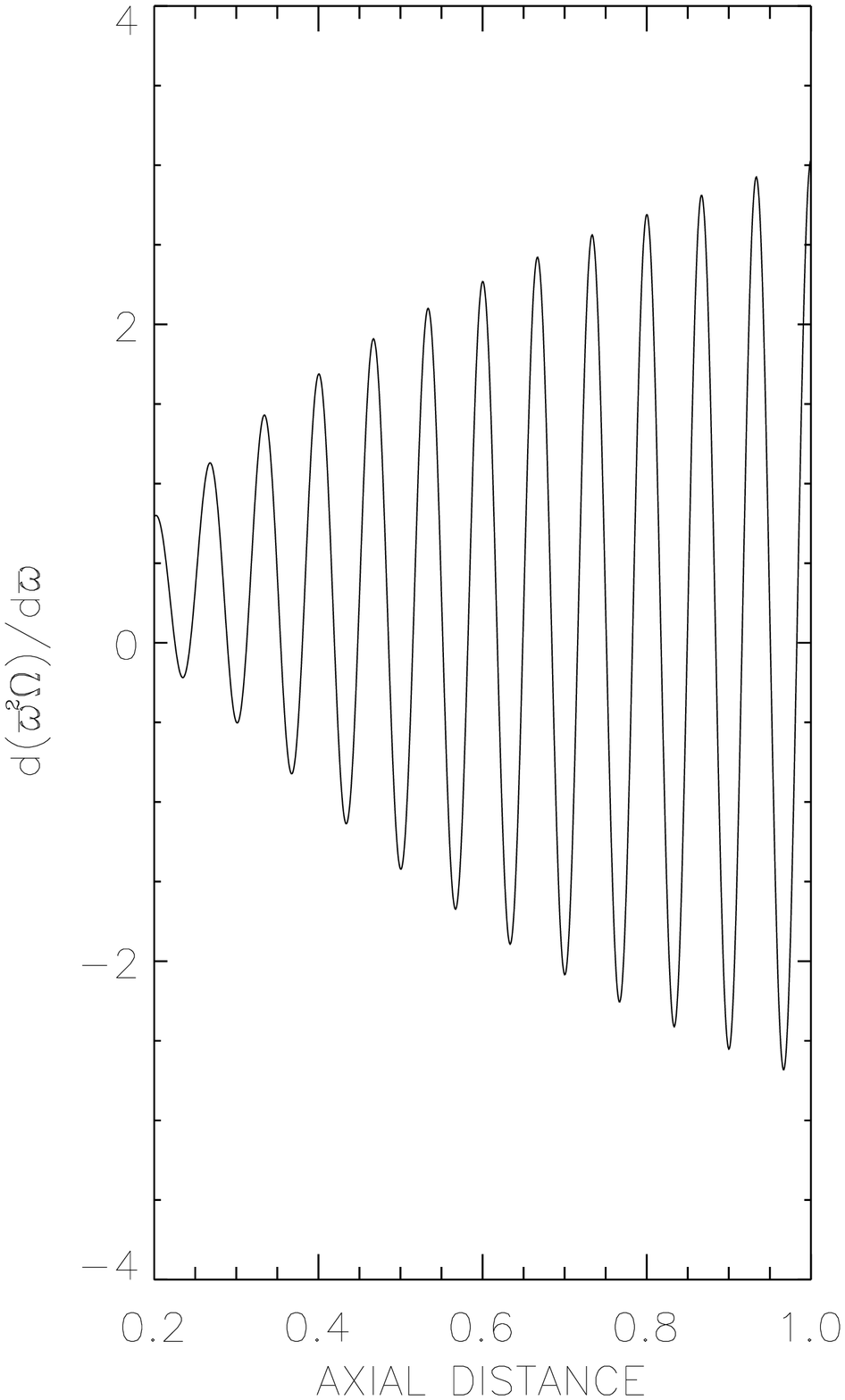,width=10cm,height=7.0cm}}
\caption{LEFT: The mean Keplerian shear flow (dotted line, $\varepsilon=0$) 
is perturbed  by sine-like fluctuations. The full line in the diagram 
represents the resulting velocity profile for the values
$\varepsilon =0.125$ and $\delta=12\pi$, the latter corresponding to
approximately $12$ periods.
RIGHT: Negative values indicate local violations of the Rayleigh stability criterion
} 
\label{Ro}
\end{figure}

The mean flow is assumed to be Keplerian for large radii but
perturbed by sinusoidal variations, i.e.
\begin{equation} \vec{\Omega}=\Omega_0 \frac{\left(1+\varepsilon \sin
(2 \delta  \varpi / \varpi_0) \right)}
{\sqrt{1 + (\varpi / \varpi_0)^3}} \vec{e}_z,
\label{2}
\end{equation}
thus we are considering the noise induced by the perturbations
as being part of the equilibrium profile (see Fig. \ref{Ro}).

This rotation law depends only on the distance 
$\varpi=r \sin \theta$ to the rotation axis,
where the parameter $\varpi_0$ is given by 
$\varpi_0=(R_{\mbox{\scr o}}-R_{\mbox{\scr i}})/2$.
The perturbations we use are characterized by the
parameter $\varepsilon$ describing the amplitude
of the fluctuations and the number of periods $\delta/\pi$
placed in a half shell diameter.
This choice of localized perturbations
allows for a smooth transition to the unperturbed
Keplerian shear flow  by decreasing the amplitude $\varepsilon$.
Because of numerical requirements we have restricted ourselves to 
$\delta=12\pi$ which can give rise to small-scale structures
of radial dimensions of 1/24.

According to the Rayleigh criterion 
a purely Keplerian shear flow
is linearly stable with respect to infinitesimal axisymmetric 
perturbations, however due to fluctuations
this stability criterion can be  violated locally. 
It happens locally for all $\varepsilon$ values exceeding 0.007.\footnote{As for 
finite value of $\varepsilon$ the basic flow may have a number of inflection points it is 
also  worth to mention Rayleigh's inflection point theorem after which it is a 
{\em necessary} condition for instability to infinitesimal disturbances (for 
parallel shear flows) that an inflection point exists. But there is no claim 
that any velocity profile with an inflection point is unstable (Acheson 1990).}

Since this instability does not depend on the behavior of
temperature, pressure and density
we do not have to take into account the temperature equation.
Moreover, for simplicity we assume the density to be constant
throughout the whole shell.

Length and time are normalized with respect to the
difference in radii and to the viscous diffusion time $\tau = (R_{\rm
o} - R_{\rm i})^2/\nu$, i.e.
\begin{equation}
\vec{r}=(R_{\mbox{\scr o}}-R_{\mbox{\scr i}}) \hat{\vec{r}}, \quad 
t =  \frac{(R_{\mbox{\scr o}}-R_{\mbox{\scr i}})^2}{\nu} \hat{t}.  
\label{3}
\end{equation}
Then the velocity field is normalized by the
rate $\Omega_0$ of the prescribed differential rotation,
\begin{equation}
\vec{u} = (R_{\mbox{\scr o}}-R_{\mbox{\scr i}}) \Omega_0 \hat{\vec{u}}.
\label{4}
\end{equation}
Expressed in terms of these variables 
the Navier-Stokes equation for the fluctuations 
$\per{\vec{u}}=\vec{u}-\vec{u}_0$ becomes
\beg 
\frac{\partial \per{\vec{u}}}{\partial t} - \Delta \per{\vec{u}}= - 
 \grad p  
+ \mbox{Re}\left(\per{\vec{u}} \times  \rot \per{\vec{u}} 
+ \vec{u}_0 \times  \rot \per{\vec{u}}  
+ \per{\vec{u}} \times  \rot \vec{u}_0\right)
\label{NS}
\ende 
(the hats are now omitted) 
with the Reynolds number
\begin{equation} \label{parameter}
\mbox{Re}
=\frac{\Omega_0 (R_{\mbox{\scr o}}-R_{\mbox{\scr i}})^2}{\nu} .
\end{equation}
Since the vector $\vec{u}$  is divergence-free 
it can be represented by
toroidal and poloidal components (Chandrasekhar 1961)
\begin{equation} 
\per{\vec{u}} = \vec{r} \times \grad \Bigl(\frac{\Phi}{r}\Bigr) + 
\rot  \Bigl(\vec{r} \times \grad
\bigl(\frac{\Psi}{r}\bigr)
\Bigr).
\label{5}
\end{equation}

We consider stress-free boundary conditions for
the flow. Thus the cross-components 
$\pi_{r \phi}$, $\pi_{r \theta}$ of the
viscous stress-tensor $\pi_{ij}=-\rho \nu
(\per{u_{i,j}}+\per{u_{j,i}})$ 
have to vanish, which can be expressed by the scalar potentials $\Psi$ and
$\Phi$ as 
\begin{eqnarray}
\frac{2}{r}\frac{\partial \Psi_{lm}}{\partial r} -
\frac{\partial ^2 \Psi_{lm} }{\partial ^2 r} = 0 , \quad
\frac{\partial}{\partial r}\Bigl(\frac{\Phi_{lm}}{r^2}\Bigr) = 0. 
\label{6}
\end{eqnarray}
No normal flow is allowed at
$r=r_{\mbox{\scr i}},r_{\mbox{\scr o}}$ hence the potential 
$\Psi$ vanishes at the boundary,
\begin{equation}
 \Psi_{lm} = 0.
\end{equation} 
\section{Numerics}
The numerical simulations have been performed by the hydrodynamical
part of the spectral magne\-to-convection code which is described in 
full detail in Hollerbach (2000). 
The spectral expansion uses spherical harmonics on the 2-sphere
and Chebyshev polynomials in the radial direction
\begin{equation}
\Psi=\sum_{k,l,m} \Psi(k,l,m) T_{k-1}(x) P^{|m|}_l(\cos \theta )e^{im\phi}
\label{7}
\end{equation} 
with the radius $r$ mapped to the interval $\lbrack -1,1 \rbrack$ via
\begin{equation}
r=\frac{r_{\mbox{\scr o}}+r_{\mbox{\scr i}}}{2}+\frac{r_{\mbox{\scr o}}-
r_{\mbox{\scr i}}}{2}x.
\label{8}
\end{equation}
The radius ratio has been fixed to 1/5 throughout while
setting the shell diameter to unity: $r_{\rm o}-r_{\rm i}=1$. In all
diagrams with a radial dependence we show plots versus the ratio
$r/r_{\rm o}$, normalizing the outer radius to unity. 

The diffusive terms in combination with the boundary conditions
are treated implicitly in spectral space,
whereas the nonlinear terms  are evaluated explicitly on a grid of 
collocation points. The timestepping is performed by  a modified 
second order Runge-Kutta scheme: first a predictor step
calculates an estimated value, and afterwards the
resulting spectral coefficients are used for evaluating again
the velocity field, which yields a corrected value in a second step. 

Because of memory requirements we had to run these simulations
on a Convex HPP-1200, where a single time step for a numerical
resolution of 40 Chebyshev-, 70 Legendre polynomials  and  3
Fourier modes required approximately 5 seconds of CPU time.
At Reynolds numbers of $\mbox{Re}=3000-5000$ the code
runs stably with a time-step of $3 \times 10^{-5}$. 
Thus a run simulating the flow over a period of half a viscous
time can be performed within one day of CPU time.
In order to avoid aliasing effects when transforming from
real space to spectral space and vv. we used a collocation
grid of the following dimension: 
$\dim (r_i, \theta_j, \phi_k)=80 \times 105 \times 5$.
\section{Results and discussion}
We have run simulations in the range Re$=3000-5000$
with different values for the amplitude $\varepsilon$ so that the 
corresponding minimum perturbation can be 
determined.  A single run where the instability leads to
a nondecaying state is discussed 
in full detail in the following subsection.
\subsection{The exemplary case Re$=3500$}
\begin{figure} 
\resizebox{100mm}{!}{\includegraphics[]{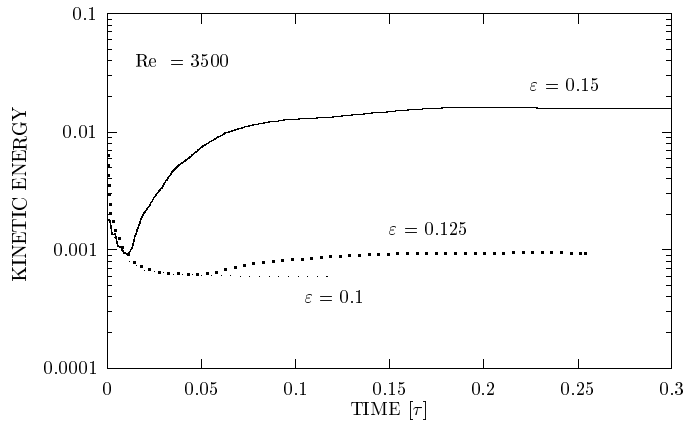}}
\caption{The kinetic energy reaches a turbulent equilibrium
for $\varepsilon=0.15, 0.125$, but decays for $\varepsilon=0.1$.
The case $\varepsilon=0.125$ is close
to the critical amplitude.  Time is measured in units of the 
viscous time.}
\label{energy}
\end{figure}

Figure \ref{energy} shows the temporal evolution of the total kinetic 
energy for a run with Re$=3500$ and several amplitudes 
$\varepsilon=0.1,\;0.125,\; 0.15$ starting from arbitrary initial fields. 
For $\varepsilon=0.15$ the solution grows by one
order of magnitude during a time of about $10\%$ of a viscous time 
after which a turbulent state has been reached.
The largest contribution to the kinetic
energy is due to the azimuthal component, whereas the radial
and meridional components yield contributions which are both smaller
by almost two orders of magnitude.
A perturbation with $\varepsilon=0.125$
still leads to a turbulent configuration, but it is close to
the critical amplitude since the kinetic energy decays for
$\varepsilon=0.1$. 

\begin{figure} 
\resizebox{100mm}{!}{\includegraphics[]{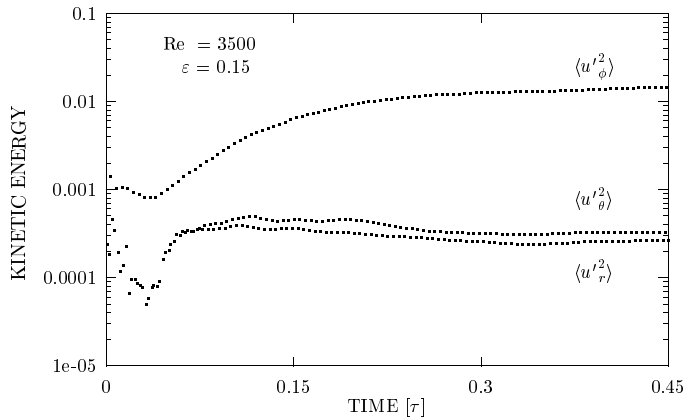}}
\caption{The kinetic energy of the turbulent state is dominated by
the azimuthal component, whereas the other components are
smaller by two orders.}
\label{energy3}
\end{figure}

Figure \ref{energy3} demonstrates the
dominant role of the azimuthal velocity fluctuations in the
small-scale flow-pattern. The turbulence field is highly
anisotropic with a massive ``Austausch'' in the $\phi$-direction.
As already stressed by Biermann (1951) such a flow field
contributes to a positive correlation $\langle u_r u_\phi\rangle$ under
the influence of basic rotation (cf. R\"udiger 1989, p. 254) --
in contrast to convective patterns, where negative
correlations do appear.

As the next step we discuss the spectral distribution of energy.
The resulting flow pattern is nearly axisymmetric and steady,
higher modes are suppressed by several orders of magnitude:
$E_0:E_1:E_2= 1: 10 ^{-2}: 10^{-4}$.
Simulations with higher azimuthal resolution
lead to the same quantitative result. For stronger amplitudes
$\varepsilon$ the non-axisymmetric modes become much stronger, but
in addition to the numerical difficulties that we encountered in simulating
those flows such strong perturbations seem to be physically irrelevant.

\begin{figure} 
\resizebox{100mm}{!}{\includegraphics[]{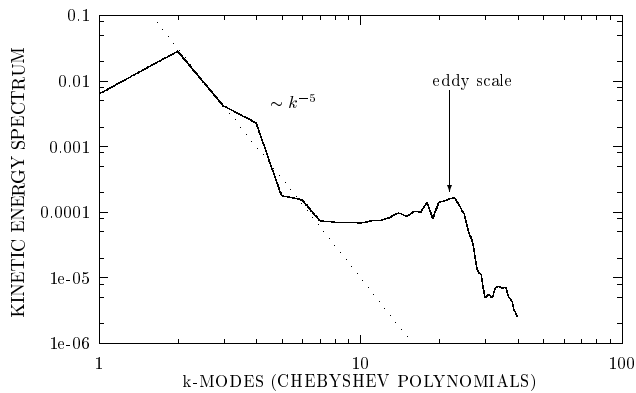}}
\caption{The Chebyshev spectrum of the kinetic energy for Re$=3500$
and $\varepsilon=0.15$ decreases like $k^{-5}$ for large modes. 
The peak at $k=22$ corresponds to the
radial dimension $\ell_{\rm corr}=0.0\overline{45}$ of a single eddy.}
\label{chebyshev}
\end{figure}

The Chebyshev spectrum (Fig. \ref{chebyshev}) behaves like $k^{-5}$ for small $k$  and 
shows strong peaks at $k_1=1$ and $k_2=22$,
the former corresponding to the global torus-like structure of the flow.
The latter peak determines the radial dimension of the turbulence elements
$\ell_{\rm corr}\simeq 1/k_2=0.0\overline{45}$.

\begin{figure} 
\resizebox{100mm}{!}{\includegraphics[]
{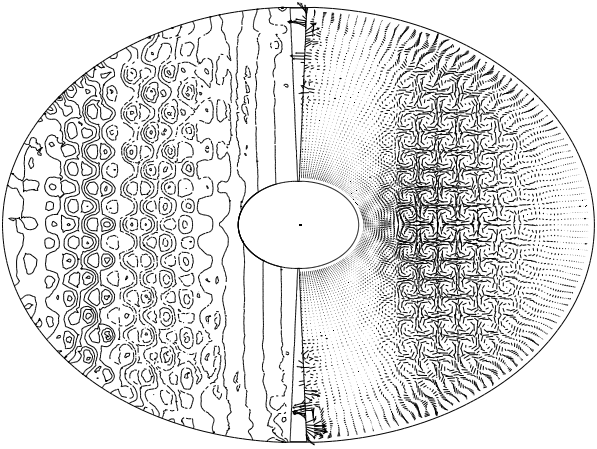}}
\caption{Left: Contour plot of the azimuthal velocity component. Right:
Meridional structure  
of the velocity field. Re$=3500$, $\;\varepsilon=0.15$. 
The eddies form
Taylor-Proudman columns 
in the range $0.4 < \varpi < 0.8$.}
\label{9}
\end{figure}

As can be seen from the energy spectrum for the
Legendre polynomials the flow is symmetric with respect to the
equatorial plane. In general, even $l$-modes are stronger
by one order of magnitude with a maximum at $l=46$ 
in comparison with odd modes resulting in a total parity of 
0.98. 
The flow is mainly confined to the region $0.4 < \varpi < 0.8$
where the eddies form Taylor columns. The alignment
which is to be expected for high Taylor numbers can also be
seen from the contour plot of the radial flow component.
We find 9 eddies in the radial and up to
18 eddies in the vertical direction depending on the height of the
shell (Fig. \ref{9}).

Despite the fact that the total flow looks rather inhomogeneous
from a global point of view we make some rough estimates for its
eddy viscosity which will agree approximately with the 
viscosity derived later on.
 From dimensional analysis one is led to 
\begin{equation}
{\nu_{\rm T}\over \nu} =u_{\rm T} \cdot \ell_{\rm corr}, 
\end{equation}
where $\ell_{\rm corr}$ and $u_{\rm T}$ denote a characteristic radial dimension
and velocity. If we insert in this relation a typical
eddy velocity  and the shell diameter we end up with a rather
small value of $\nu_{\rm T} / \nu \approx 0.03$.

\begin{figure} 
\resizebox{100mm}{!}{\includegraphics[]{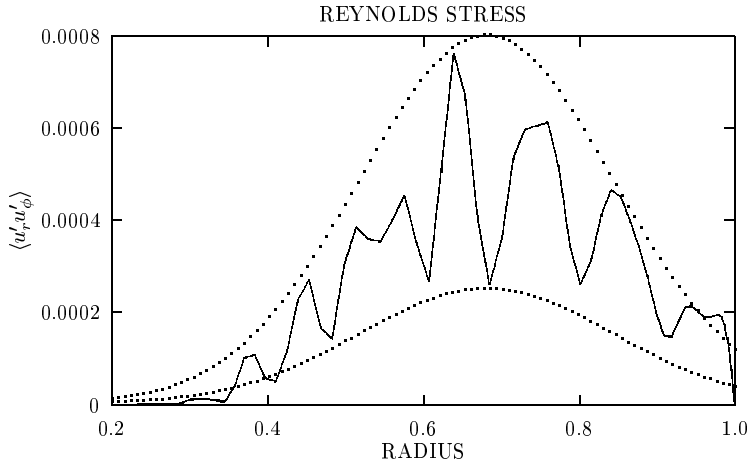}}
\caption{The cross correlation of the velocity field has a Gauss-type
structure in radius. The dotted lines show analytical curves 
$\exp {-(\varpi-\varpi_0)^2/(2\sigma^2)}$ used for
fitting the minima and maxima: $\varpi_0\approx0.68$, $\sigma\approx
0.204$.   $\varepsilon=0.15$, Re $ =3500$.}
\label{10}
\end{figure}
\begin{figure} 
\resizebox{100mm}{!}{\includegraphics[]{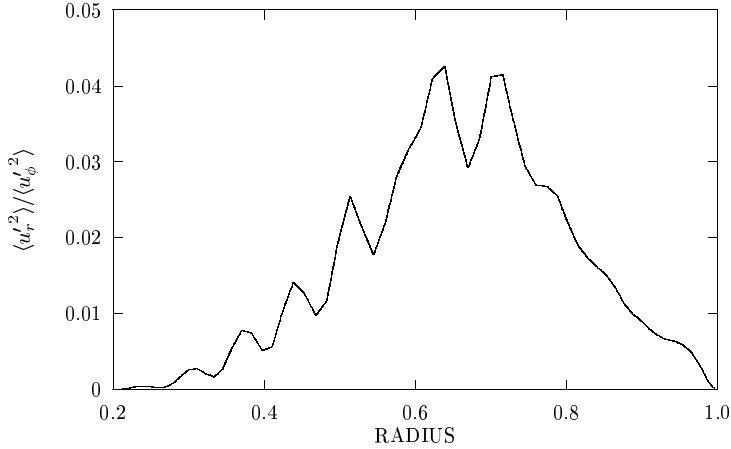}}
\caption{The intensity ratio in the small-scale flow pattern of the
model corresponding to Fig. 6. The azimuthal perturbations dominate 
the radial ones everywhere in opposition to the condition (1) valid 
for convection.}
\label{11}
\end{figure}
Now we turn to the angular momentum transport
which is regulated by the dissipative stress.
Figure \ref{10} shows the Reynolds stress $\langle \per{u}_r
\per{u}_{\phi}\rangle$ as a function of the radius calculated by integrating
over the 2-sphere for the amplitude $\varepsilon=0.15$. Strong
correlation of the velocity field can be found only in the region
$ 0.5 < r < 0.9 $, the fluctuations being exponentially damped
outside, but they are positive in the entire computational
domain. Indeed after Fig. \ref{11} there is no exception from
the dominance of $\langle u_\phi'^2\rangle$ in striking
difference to relation (\ref{0}) valid for convective Kepler disks.

\begin{figure} 
\resizebox{100mm}{!}{\includegraphics[]{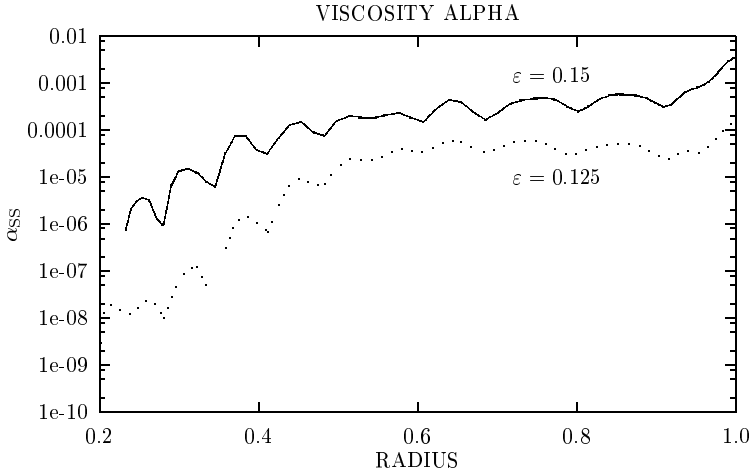}}
\caption{The viscosity $\alpha_{\mbox{\scr
SS}}$ is positive and nearly constant between $0.5<r<0.9$.
The value of  $\alpha_{\mbox{\scriptsize SS}}$ differs by one order
for the amplitudes $\varepsilon=0.15$ and  $\varepsilon=0.125$. }
\label{12}
\end{figure}

The maxima and minima are forming a Gaussian,
\begin{equation}
\mbox{extremum}\langle \per{u_r} \per{u_\phi} \rangle 
\propto \exp (-12(r-0.68)^2),
\label{13}
\end{equation}
with a standard deviation of about $\sigma \approx 0.204$.
We proceed to derive the eddy viscosity $\nu_{\rm T}$  
by integrating (\ref{13}) with respect to the radius
 and normalizing it with the unperturbed Keplerian shear flow 
$\Omega_{\rm Kep}$:
\begin{equation} \label{veddy}
\frac{\displaystyle \nu_{\rm T}}{\displaystyle \nu}
{\varpi \left( \frac{d\Omega_{\rm Kep}}{\displaystyle d\varpi}\right)}
=-\mbox{Re} \;\langle \per{u_r} \per{u_{\phi}} \rangle .
\end{equation}
This prescription yields in the case Re $=3500$ 
for the ratio of turbulent and molecular viscosity
$\nu_{\rm T}/\nu= 0.15$. 
This turbulent viscosity will now be related to the characteristic length
and velocity.
Shakura \& Sunyaev (1973) have introduced
a viscosity alpha  
using the scale height
$H(\varpi)=\sqrt{r_o^2-\varpi^2}$.
All uncertainties about the turbulence has been put into
this single parameter which allows to construct a variety
of $\alpha$-type models in accretion disk theory,
\begin{equation}
\nu_{\rm T}=\alpha_{\mbox{\scriptsize SS}}\,\Omega_{\rm Kep} H^2.
\end{equation}
The resulting radial structure for $\alpha_{\mbox{\scriptsize SS}}$ 
is shown in Fig. \ref{12} for
$\varepsilon=0.15$ and $\varepsilon=0.125$. It
is positive everywhere. A
complete volume integration yields the value 
$\alpha_{\mbox{\scriptsize SS}}= 3.75 \times 10^{-4}$.
In the ``turbulent'' region $0.5 < r < 0.9$ the viscosity alpha is nearly
constant, while the value for $\varepsilon=0.125$ is
by one order smaller than for $\varepsilon=0.15$.
Those low values for all viscosity parameters indicate that
we have not found a truly turbulent state, which is confirmed
by the fact that the velocity field is stationary.  In the next
section we study the question whether this state can be
sustained for a purely Keplerian rotation law.  
\subsection{Perturbations as initial fields}
After the  Rayleigh criterion  a pure Keplerian shear flow
is stable against axisymmetric, infinitesimal perturbations,
but it does not make predictions about the stability of
finite amplitude perturbations.
In order to check whether  the nonlinear turbulent state 
discussed in the foregoing subsection can give rise to such
a nonlinear instability the resulting flow was used as initial field 
for a simulation when the perturbations have been switched off,
i.e. $\varepsilon=0$.  Notice that the
initial state contains weak non-axisymmetric contributions.

\begin{figure} 
\resizebox{100mm}{!}{\includegraphics[]{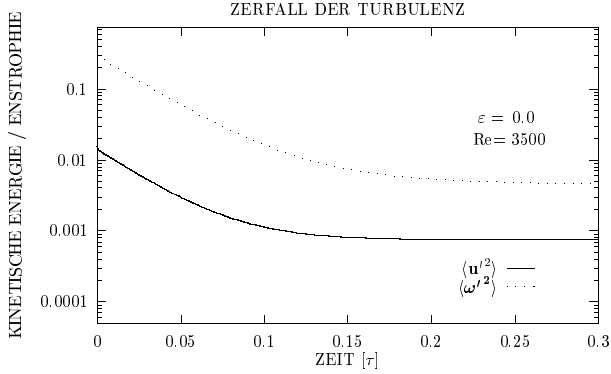}}
\caption{The kinetic energy $\langle \per{\vec{u}}^2\rangle$ and the
enstrophy $\langle \per{\mbox{\boldmath$\omega$}}^2\rangle$ decay 
exponentially within
$t=0.1\, \tau$ as soon as the perturbations $\varepsilon$ are switched off.}
\label{14}
\end{figure}

It turns out that the kinetic energy $\langle \per{\vec{u}}^2\rangle$
and the enstrophy $\langle \per{\vec{\mbox{\boldmath$\omega$}}}^2\rangle$  
decay exponentially
very fast within a time of 0.15 diffusion times, $\tau$,
corresponding to approximately 525 orbits (Fig. \ref{14}). 
Within a few orbits the eddy viscosity
decreases strongly since the rotational support from the
sinusoidal perturbations is missing. After this short period  it has 
reached approximately the value predicted by our estimate of the
foregoing section using dimensional
analysis: $\nu_{\rm T} / \nu \approx 0.02$. Then it decays like
$\nu_{\rm T} / \nu \sim e^{-27 t/\tau}$.

Both the enstrophy and the energy dissipation
rate $d \langle \per{\vec{u}}^2 \rangle /dt$ are used for a
comparison with the decay of homogeneous turbulence 
by rewriting the energy balance equation
\begin{equation}
\frac{\nu_{\mbox{eff}}}{\nu}=
-\frac{1}{2}\frac{\frac{\displaystyle d}{\displaystyle dt}
\langle \per{\vec{u}}^2\rangle}
{\langle \per{\mbox{\boldmath$\omega$}}^2\rangle} ,
\end{equation}
where  $\nu_{\mbox{eff}}$ accounts for the effective nonlinear contributions 
stemming from the energy flux and for the energy injection by the forced
differential rotation. 
As long as the flow shows strong local inhomogeneities this value
remains nearly constant, but after $t=0.1 \,\tau$, when the flow
consists of several almost homogeneous domains, an exponential
decay like $\nu_{\mbox{eff}}/\nu\sim e^{-33 t/\tau}$ sets in.
Hence,  the turbulence does not survive the transition to a   pure, undisturbed
Kepler flow.
\section{Summary and conclusions}
The results for runs with several Reynolds numbers and with different
perturbation amplitudes are summarized in Table 1 starting with
Re$ =3000$.

Since for stress-free boundary conditions
the angular momentum is conserved, the 
kinetic energy does not decay  to zero completely even for a 
purely Keplerian shear flow if the initial fields contain nonvanishing 
angular momentum. Therefore we started our series of runs at
Re $=3000$ without any perturbation. 
A calculation of the angular
momentum yielded $L_z \approx -5.4 \times 10^{-2}$ in this case. We are
considering the corresponding kinetic energy $E_{\mbox{kin}} \approx
-5.89 \times 10^{-4}$ as an offset for the subsequent runs with 
perturbations to the velocity profile $\varepsilon \not= 0$ rather
than performing a transformation such that $L_z=0$.

The minimum value of the perturbation amplitude for reaching a
turbulent state at Re$ =3000$ turned out to be 
$\varepsilon=0.15$. The corresponding kinetic energy exceeded
the energy offset due to initial angular momentum just a little
bit, thus indicating that this case is close to neutral stability.
The eddy viscosity accepted already a steady finite value of
order $10^{-2}$ while the viscosity alpha reached approximately $10^{-5}$.
Increasing the amplitude to $\varepsilon=0.175$ all values
grow by approximately one order of magnitude. 
Even for strong perturbations we never find a viscosity alpha 
larger than $10^{-4}$ indicating, therefore, no efficient mechanism of 
angular momentum transport in this model.
The kinetic energy has been vastly dominated in all runs by the azimuthal
component throughout.

The onset of turbulence is mainly governed by the Reynolds number,
for which one usually assumes that the critical amplitude 
behaves according to a power law like 
$\varepsilon_{\rm crit} \sim  \mbox{Re}^{-\beta}$. 
For plane Couette flow an analytical treatment by
Dubrulle \& Zahn (1991) yielded for this parameter $\beta =1/3$.

Using the critical amplitudes which we have determined in our runs 
 we find roughly the following scaling
\begin{equation}
\varepsilon_{\rm crit} \sim \mbox{Re}^{-1}.
\end{equation}
This behaviour predicts that the instability will set in at high
Reynolds numbers already for weak perturbations, but
as our simulations have shown the nonaxisymmetric modes 
(being necessary for the occurrence of true turbulence)
will be excited only for fairly strong perturbations.
Accordingly, the turbulence was not sustained when the
perturbations have been switched off, instead the eddy structures
decayed very fast. 
Thus, our studies give no indications that 
Kepler rotation is nonlinearly unstable against finite
amplitude perturbations (for 
Reynolds numbers up to $10^3$).

\begin{table}[]
\caption{The total kinetic energy, the eddy viscosity and the viscosity alpha 
for several combinations of Reynolds numbers and perturbation amplitudes.}

\vspace*{0.5cm}
\begin{tabular}{|c|l|c|c|c|}
\hline
Re& $\varepsilon$ & $\langle\vec u'^2\rangle$ & $\nu_{\rm T}/\nu$
& $\alpha_{\rm SS}$\\
\hline
&&&&\\[0.5ex]
3000 & 0.0 & 5.89 $\times 10^{-4}$ & $\approx 10^{-5}$ & $\approx 10^{-8}$\\
     & 0.125 & 5.97 $\times 10^{-4}$ & $\approx 10^{-5}$ & $\approx 10^{-8}$\\
     & 0.15  & 7.34 $\times 10^{-4}$ & 9.45 $\times 10^{-3}$ & 1.25 
     $\times 10^{-5}$\\ 
     & 0.175 & 8.55 $\times 10^{-3}$ & 1.03 $\times 10^{-1}$ & 1.84 $\times 
     10^{-4}$\\
3500 & 0.1 & 5.95 $\times 10^{-4}$ & $\approx 10^{-5}$ & $\approx 10^{-8}$\\
     & 0.125 & 9.37 $\times 10^{-4}$ & 1.72 $\times 10^{-2}$ & 
     2.75 $\times 10^{-5}$\\  
     & 0.15 & 1.56 $\times 10^{-2}$ & 1.50 $\times 10^{-1}$ & 
     3.75 $\times 10^{-4}$\\
4000 & 0.1 & 5.85 $\times 10^{-4}$ & $\approx 10^{-5}$ & $\approx 10^{-8}$\\
     & 0.11 & 1.73 $\times 10^{-3}$ & 3.70 $\times 10^{-2}$ & 
     6.63 $\times 10^{-5}$\\
5000 & 0.075 & 5.74 $\times 10^{-4}$ & $\approx 10^{-6}$ & $\approx 10^{-9}$\\
     & 0.085 & 1.72 $\times 10^{-3}$ & 4.2 $\times 10^{-2}$ & 7.52 $\times 
     10^{-5}$\\
     & 0.09 & 3.84 $\times 10^{-3}$ & 7.71 $\times 10^{-2}$ &
      1.53 $\times 10^{-4}$\\
 \hline
\end{tabular}
\end{table}


\addresses
\rf{G\"unther R\"udiger,
Astrophysikalisches Institut Potsdam,
An der Sternwarte 16,
D-14482 Potsdam,
Germany,
e-mail: GRuediger@aip.de}
\rf{Andreas Drecker, Astrophysikalisches Institut Potsdam, 
An der Sternwarte 16, D-14482 Potsdam, Germany
     } 

\end{document}